\documentclass[aps,pra,twocolumn,floatfix,superscriptaddress,nobalancelastpage]{revtex4-1}

\usepackage[utf8]{inputenc}
\usepackage[english]{babel}
\usepackage{amsfonts}
\usepackage{amsmath}
\usepackage{graphicx}
\usepackage{mathtools}
\usepackage[caption=false]{subfig}

\usepackage{soul}
\usepackage{color}

\newcommand{\eps}{\varepsilon}

\newcommand{\cose}[1]{\mathrm{cos}\left(#1\right)}

\newcommand{\daga}[1]{#1^{\dagger}}
\newcommand{\ex}[1]{\:e^{#1}}
\newcommand{\Ket}[1]{\left|#1\right>}
\newcommand{\Bra}[1]{\left<#1\right|}
\newcommand{\BraKet}[2]{\left<#1|#2\right>}

\newcommand{\trace}[1]{\mathrm{tr}\left(#1\right)}
\newcommand{\ad}[1]{\mathrm{ad}_{#1}}
\newcommand{\adc}[1]{\mathrm{ad}^2_{#1}}
\newcommand{\conmu}[2]{\left[#1,#2\right]}

\begin{document}

\title{Geometric quantum speed limits and short-time accessibility to unitary operations }

\author{Pablo M. Poggi}
\email{ppoggi@unm.edu}
\affiliation{Center for Quantum Information and Control, University of New Mexico, MSC07-4220, Albuquerque, New Mexico 87131-0001, USA}
\date{\today}


\begin{abstract}
	The notion of quantum speed limit (QSL) refers to the fundamental fact that two quantum states become completely distinguishable upon dynamical evolution only after a finite amount time, called the QSL time. A different, but related concept is that of minimum control time (MCT), which is the minimum evolution time needed for a state to be driven (by suitable, generally time-dependent, control fields) to a given target state. While the QSL can give information about the MCT, it usually imposes little restrictions to it, and is thus unpractical for control purposes. In this work we revisit this issue by first presenting a theory of geometrical QSL for unitary transformations, rather than for states, and discuss its implications and limitations. Then, we propose a framework for bounding the MCT for realizing unitary transformations that goes beyond the QSL results and gives much more meaningful information to understand the controlled dynamics of the system at short times.
	
\end{abstract}

\maketitle

\section{Introduction}

Realizing the prospect of quantum-enabled technologies, such as computation and simulation, demands an extremely high degree of precision in the control of quantum systems. Extraordinary advances have been made over the past years on increasing the fidelities for one- and two-qubit operations in various quantum computing platforms \cite{monroe_exp2017,molmer2017,watson2018}, and on precise engineering of interactions in synthetic quantum materials \cite{monroe2017,bloch2017}. Since any unitary operation is ultimately the result of a continuous-time evolution of a quantum system, to avoid errors the goal is to control these systems as fast and accurately as possible, in order to reduce the non-unitary effects induced by coupling to the environment. Consequently, it is important to understand the scope and limitations of controlled quantum dynamics at very short times.\\

One fundamental limitation is imposed by the so-called quantum speed limit (QSL). In its original formulation \cite{fleming1973,bhatta1983,pfeifer1993}, the QSL referred to the fact that the state of a quantum system evolving according to a time-independent Hamiltonian would require a finite amount of time to become completely distinguishable (i.e. orthogonal) to itself. Later on, this notion was developed as a geometrical result generalizing to the case of non-orthogonal states \cite{giovannetti2003,levitin2009,pires2016}, leading up to results for driven \cite{deffner2013_teur} and open quantum systems \cite{delcampo2013,deffner2013,taddei2013} (a recent review article on this topic can be found on \cite{deffner2017}).\\

From the perspective of quantum control, the geometric QSL is a fundamental limitation to how fast we can achieve a predefined target state or transformation. However, a more stringent limitation is given by the fact that, in every case of interest, we have a limited set of controls available. Note that, under certain assumptions, one can show that a system is fully controllable \cite{dalessandro}, meaning that by suitably changing in time that limited set of control fields one can perform any unitary operation in the corresponding group, say $SU(d)$ for a $d$-dimensional Hilbert space. Finding the particular shape of those fields is usually done using the tools of quantum optimal control (QOC), which have proven to be extremely useful and versatile in various quantum information platforms over the past two decades \cite{gambetta2010,doria2011,anderson2015,glaser2015}.\\

In QOC problems, we consider a driving Hamiltonian $H[\epsilon(t)]$, where $\epsilon(t)$ represents some time-dependent control field (or a set of them). Given a target unitary $V$ and the evolution time $T$, we aim to minimize $||U(T)-V||$ in some appropriate metric, where $U(T)$ is the evolution operator at the final time. It is expected that for each particular target, there is a minimum control time (MCT), i.e., the minimum value of the evolution time $T$ for which the optimization has (in principle) a solution. For any given target $V$, the MCT will typically be larger than the QSL time \cite{poggi2013}. Obtaining the MCT is, however, not an easy task. One approach consists in performing a two-objective optimization to reduce the error in implementing the target, while simultaneously shortening the protocol as much as possible \cite{carlini2007,lapert2012,wang2015}. An alternative is to look for solutions within the usual QOC scheme for different values of $T$ \cite{caneva2009,tibbetts2012,poggi2015}. In both cases, the optimization becomes increasingly difficult and computationally expensive \cite{sorensen2016,larocca2018}. It is therefore desirable to establish a framework that allows us to obtain bounds or estimates on the minimum time required to implement a predefined gate. Going beyond the geometrical QSL, these bounds should take into account a given set of resources, related to which elements of the generating algebra we can manipulate in time.\\

%

In this paper we first explore geometric QSL bounds in the space of unitary operations, and discuss their application to quantum control problems. We do a thorough exploration of the MCT for $SU(2)$ and $SU(3)$ models in order to gain insight on how the available controls in the generating Hamiltonian restrict the unitaries that can be achieved at a given time. Also, by studying the short-time behavior of the time-dependent Schr\"odinger equation for arbitrary control fields, we obtain bounds on the MCT that are much more restrictive than the geometric QSL.\\

This paper is organized as follows. In Section \ref{sec:geom_qsl} we revisit the geometric formulation of the QSL for states, and then develop an analogous construction for unitary transformations. We derive two families of bounds and give examples for $SU(2)$ and $SU(3)$. In Section \ref{sec:control} we study the MCT for these examples, and present an analysis of the short-time behavior of quantum systems driven time-dependent fields. This will allow us to construct bounds on the MCT that will serve as a refinement of the QSL results. Finally in Section \ref{sec:conclusions} we discuss potential future directions of work on these topics.\\

\section{Geometrical Quantum Speed Limit for unitary operations} \label{sec:geom_qsl}
\subsection{QSL for pure state evolution}
\begin{figure}
	\includegraphics[width=0.8\linewidth]{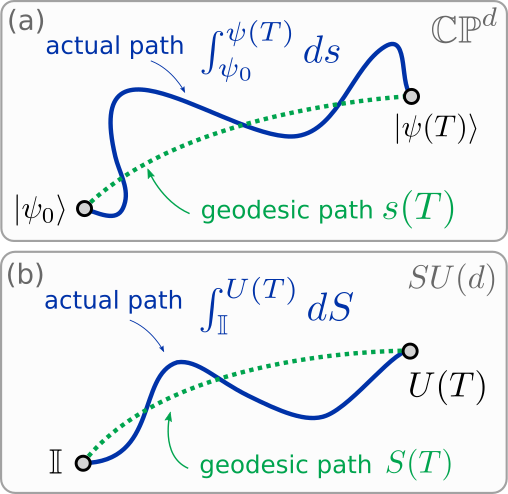}
	\caption{\label{fig:fig1} Schematic picture of the evolution of a quantum system. (a) paths between an initial state $\Ket{\psi(t=0)}=\Ket{\psi_0}$ and a final state $\Ket{\psi(T)}$ in the complex projective space $\mathbb{CP}^d$ associated with a $d$-dimensional Hilbert space. (b) paths between unitary transformations between the initial point $U(t=0)=\mathbb{I}$ and final point $U(T)$ in $SU(d)$.}
\end{figure}

Let us first briefly revisit the QSL formulation for (pure) state evolution. For that, we recall the definition of the Fubini-Study distance between two states $\Ket{\psi_1}$ and $\Ket{\psi_2}$ \cite{bengtsson2017},
\begin{equation}
s\left(\psi_1,\psi_2\right)=2\arccos\left(\lvert \BraKet{\psi_1}{\psi_2}\rvert \right)
\end{equation}
Take $\Ket{\psi(t)}$ to be the state of a $d-$dimensional quantum system that evolves according to $i\hbar\frac{d}{dt}\Ket{\psi(t)}=H(t)\Ket{\psi(t)}$. Consider a curve $\mathcal{C}$ in the complex projective space $\mathbb{CP}^d$, given by the set of points $\left\{\Ket{\psi(t)}\right\}_{0\leq t\leq T}$. The length of $\mathcal{C}$ according to the Fubini-Study metric was derived by Anandan and Aharonov \cite{anandan1990}, yielding
\begin{equation}
\mathrm{length}(\mathcal{C})=\frac{2}{\hbar}\int_0^T \Delta E(t)\:dt,
\end{equation}
\noindent where $\Delta E(t)=\sqrt{\langle H(t)^2\rangle-\langle H(t)\rangle^2}$ is taken over the state $\Ket{\psi(t)}$.
If we look at Fig. \ref{fig:fig1}, we can make the straightforward observation that the length of $\mathcal{C}$ is always smaller or equal than $s(T)\equiv s\left(\psi_0,\psi(T)\right)$, which is the length of the geodesic path connecting those two points, and where we have set $\Ket{\psi_0}=\Ket{\psi(0)}$. So, we have
\begin{equation}
s(T)\equiv s\left(\psi_0,\psi(T)\right)\leq \frac{2}{\hbar}\int_0^T \Delta E(t)\:dt,
\label{ec:qsl_st_aa}
\end{equation}
\noindent where the equality holds if and only if the system evolves along the geodesic path. Eqn. (\ref{ec:qsl_st_aa}) is usually referred to as the Anandan-Aharonov relation. From it, we can see that
\begin{equation}
T\geq \frac{\hbar\: s(T)}{2\overline{\Delta E}},
\label{ec:qsl_st_mt}
\end{equation}
\noindent where, formally, $\overline{\Delta E}$ can be regarded as the mean value of the function $\Delta E(t)$. Note, however, that such mean value will depend on the specific time-dependence of the Hamiltonian, which could be a priori arbitrary. This is especially important in control problems, where we seek a bound on the evolution time, before solving the optimization problem. The right hand side of Eqn. (\ref{ec:qsl_st_mt}) gives a lower bound on the minimum time required for state $\Ket{\psi(t)}$ to become distinguishable from $\Ket{\psi_0}$ by an amount $s(T)$ (in the Fubini-Study metric). Note that, in particular, it is also a bound for the minimum time that it should take for the system to evolve $\Ket{\psi_0}$ into any target state $\Ket{\psi_G}$ that is at a distance $s(T)$ from it.\\

Finally, note that from Eqn. (\ref{ec:qsl_st_aa}) we can recover the known Mandelstam-Tamm relation \cite{mand_tamm1945,bhatta1983} by setting $s(T)=\pi$ (its maximum possible value) and considering a time-independent Hamiltonian, thus obtaining $T\geq \frac{\pi}{2\Delta E}$.

\subsection{QSL for unitary transformations}

We now turn to deriving an expression for the QSL time for unitary operators. Some related work along this line has been reported in Refs. \cite{chau2011,lee2012}. We will proceed analogously to what was described in the previous section. Let us first define a distance between two operators $U,V\in SU(d)$ as 

\begin{eqnarray}
S_1(U,V)&=&\max\limits_{\psi}\left[ s\left(U\Ket{\psi},V\Ket{\psi}\right)\right] \nonumber \\
&=&2\:\mathrm{arccos}\left(\min\limits_{\psi}\lvert \Bra{\psi}U^\dagger V\Ket{\psi}\rvert\right) \label{ec:fs_units}
\end{eqnarray}

If we define $W=U^\dagger V$, we have that $W$ is unitary. This definition of distance was considered in Ref. \cite{acin2001}, where its shown that the optimization in Eqn. (\ref{ec:fs_units}) can be solved by defining $\delta$ as the minimum arclength such that all eigenphases of $W$, say $\left\{\varphi_k\right\}$, are included in such arc. The minimization over all states then involves two cases. If $\delta\geq\pi$, then 
\begin{equation}
\min\limits_{\psi}\lvert \Bra{\psi}U^\dagger V\Ket{\psi}\rvert = 0,
\end{equation} 

\noindent meaning that $U$ and $V$ are completely distinguishable because there exists a state $\Ket{\psi}$ such that $U\Ket{\psi}$ and $V\Ket{\psi}$ are orthogonal. If $\delta<\pi$, then such state does not exist, and the optimal case (i.e. the maximum distinguishability) is achieved by choosing $\Ket{\psi_a}=\left(\Ket{w_{min}}-\Ket{w_{min}}\right)/\sqrt{2}$ to get 
\begin{equation}
\lvert \Bra{\psi_a}U^\dagger V\Ket{\psi_a}\rvert = \cose{\frac{\delta}{2}}.
\end{equation}

Note that $\Ket{w_{max}}$ and $\Ket{w_{min}}$ are the eigenstates of $W$ corresponding to the eigenphases which are furthest apart in the circle. With this considerations we obtain a closed expression for the measure of distance $S_1$:
\begin{equation}
S_1(U,V)=\min \left(\delta,\pi\right)
\label{ec:qsl_un_dist1}
\end{equation}

Having defined a distance between unitaries, we can compute the differential element $dS_1=S_1\left(U(t),U(t+dt)\right)$. We now set $\hbar=1$ for the remainder of the paper. To compute $dS_1$ we need to look at the eigenvalues of 
\begin{equation}
U(t)^\dagger U(t+dt)=U^\dagger(t) \ex{-iH(t)dt}U(t),
\end{equation}
\noindent which are the same as those of $\ex{-iH(t)dt}$ alone. So, if we define
\begin{equation}
\Delta \varepsilon(t) = E_{max}(t)-E_{min}(t),
\label{ec:qsl_un_gen1}
\end{equation}

\noindent we have that $dS_1=\Delta \varepsilon(t) dt$. Integrating and imposing the inequality as described before, we get
\begin{equation}
S_1(T)\equiv S_1\left(U(T),\mathbb{I}\right)\leq \int_0 ^T \Delta \varepsilon(t)dt.
\label{ec:qsl_un_aa1}
\end{equation}

The inequality (\ref{ec:qsl_un_aa1}) is our first version of the the Anandan-Aharonov relation for unitary processes. Once again, if we choose some particular (time-independent) $\overline{\Delta\varepsilon}$ such that $0\leq \Delta \varepsilon(t)\leq \overline{\Delta\varepsilon}$, we can bound the integral and write
\begin{equation}
T\geq \frac{S_1(T)}{\overline{\Delta\varepsilon}}\equiv \tau_{QSL}^{(1)}.
\label{ec:qsl_un_t1}
\end{equation}

Before we proceed, let us propose an alternative way of obtaining a QSL time. For this, note that the definition in Eqn. (\ref{ec:qsl_un_dist1}) arises from optimizing the Fubini-Study distance over all possible states. The expression obtained is deceptively simple, since calculating $\delta$ requires diagonalizing $U^\dagger V$, and calculating $\Delta \varepsilon$ requires diagonalizing $H(t)$ for all times. However, we are at liberty of defining an alternative measure of distance between unitaries, which we obtain by dropping the optimization over all states in Eqn. (\ref{ec:fs_units}) and simply defining

\begin{equation}
S_2(U,V)=2\:\mathrm{arccos}\left(\frac{1}{d}\lvert \trace{U^\dagger V}\rvert \right)
\label{ec:qsl_un_dist2}
\end{equation}

We can see again that the distance depends on the overlap unitary $W=U^\dagger V$, but only via its trace. Note that this relates nicely to the usual definition of fidelity in quantum optimal control \cite{khaneja2005}. We now proceed analogously to obtain an expression for a QSL time. The differential element of distance is now
\begin{eqnarray}
\left(dS_2\right)^2 &=& 4\left(1-\frac{1}{d^2}\lvert \trace{U(t)^\dagger U(t+dt)}\rvert^2\right) \nonumber \\
&=& \frac{4 dt^2}{d}\trace{H(t)^2},
\end{eqnarray}
\noindent where we have imposed the fact that $H(t)$ is traceless and properties of the trace. Note that we can introduce $\trace{H^2}=||H||^2$ using the usual Hilbert-Schmidt inner product. We then obtain another version of the Anandan-Aharonov relation for unitaries

\begin{equation}
S_2(T)\equiv S_2(U(T),\mathbb{I})\leq \frac{2}{\sqrt{d}}\int_0^T ||H(t)||dt
\label{ec:qsl_un_aa2}
\end{equation}

From this, we can again bound the integral by using $0\leq ||H(t)||\leq \overline{||H||}$ and thus we get

\begin{equation}
T\geq \frac{\sqrt{d}}{2}\frac{S_2(T)}{\overline{||H||}}\equiv \tau_{QSL}^{(2)}
\label{ec:qsl_un_t2}
\end{equation}

The obtained expressions (\ref{ec:qsl_un_t1}) and (\ref{ec:qsl_un_t2}) give bounds $\tau_{QSL}^{i}$ ($i=1,2$) on the minimum time required for $U(t)$ to become distinguishable from $U(0)=\mathbb{I}$ by an amount $S_i(T)$ and, in particular, to reach some target transformation $V\in SU(d)$ such that $S_i(V,\mathbb{I})=S_i(T)$. Note that, in terms of control, both results for the QSL time are geometric in nature: they depend on the target process $V$ only via its distance to identity. The actual structure of the driving Hamiltonian does not come into play when computing $\tau_{QSL}^{i}$. \\

\subsection{Examples}

In this section we will use study the QSL expressions derived above in two model systems. In Section \ref{sec:control} we will come back to these models and consider a wide range of control problems which will allow us to compare the minimum control time and quantum speed limit times in an unified setting. \\

We first introduce a two-level ($d=2$) Hamiltonian 

\begin{equation}
H_2(t)=\frac{\Omega}{2}\left(\cos{\alpha(t)}\sigma_x + \sin{\alpha(t)}\sigma_y\right),
\label{ec:hami_su2}
\end{equation}

\noindent where $\left\{\sigma_i\right\}$, $i=x,y,z$ are the usual Pauli operators. This Hamiltonian describes the dynamics of a two-level atom driven by a resonant electromagnetic field with a time dependent phase $\alpha(t)$, which in principle is arbitrary \cite{scully1999}. Here $\Omega$ is the Rabi frequency, which is constant in time and sets the energy scale of the problem.\\

For $SU(2)$, it is easy to see that both metrics introduced in the previous section, c.f. eqns. (\ref{ec:qsl_un_dist1}) and (\ref{ec:qsl_un_dist2}) are equivalent, so we will drop the subindices and denote $S=S_1=S_2$. The model in Eqn. (\ref{ec:hami_su2}) has a few nice properties. First, it is fully controllable: any element of $SU(2)$ can be implemented in finite time by suitable picking the shape of the field $\alpha(t)$. Also, irrespective of the particular shape of $\alpha(t)$, the system evolves at ``constant speed'' through state space, since 

\begin{equation}
\frac{dS}{dt}=\Delta \eps=\frac{2}{\sqrt{d}}||H_2(t)||=\Omega,\: \forall\: t. 
\end{equation}

In order to evaluate the QSL bounds and to draw a connection to the control problem, we introduce a family of target unitaries

\begin{equation}
V_n(\phi)=\ex{-i \frac{\sigma_n}{2}\phi},\ 0\leq \phi\leq \pi,
\label{ec:targets_su2}
\end{equation}

\noindent where $\sigma_n=\hat{n}.\vec{\sigma}$ and $\hat{n}$ is a unit three-dimensional vector. We can calculate the distance between $V_n(\phi)$ and the identity using Eqn. (\ref{ec:qsl_un_dist2}), obtaining

\begin{equation}
S_2(V_n(\phi),\mathbb{I})=2\arccos\left(\lvert \cos{\frac{\phi}{2}}\rvert\right)=\phi.
\end{equation}

Evaluating Eqn. (\ref{ec:qsl_un_t2})  we obtain a simple QSL bound for the $SU(2)$ case

\begin{equation}
\tau_{QSL}(\phi)=\frac{\phi}{\Omega}
\label{ec:qsl_t_su2}
\end{equation}

This result gives us a very simple example of a general feature of the geometric QSL. Note that Eqn. (\ref{ec:qsl_t_su2}) is independent of $\hat{n}$. This isotropic property implies that we obtain the same bound on the time required to implement either $x$, $y$ or $z$ rotations, since all of them (for a fixed angle of rotation $\phi$) are at the same distance from $\mathbb{I}$. Note, however, that the Hamiltonian is not isotropic: $x$ and $y$ rotations are easy to perform, but $z$ rotations require more time.  We will analyze this difference in more detail in the next section.\\

Next, we introduce a three-level ($d=3$) Hamiltonian

\begin{equation}
H_3(t) = \frac{\Omega}{2}\left(\cos{\alpha(t)}\lambda_A + \sin{\alpha(t)}\lambda_B\right),
\label{ec:hami_su3}
\end{equation}

\noindent where $\lambda_A=\lambda_1$, $\lambda_B=(\lambda_2+\lambda_4)/\sqrt{2}$, and $\left\{\lambda_i\right\}$, $i=1,2,\ldots,8$ are the usual Gell-Mann matrices for $\mathfrak{su}(3)$ (i.e. the Lie algebra of skew-hermitian $3\times 3$ matrices). We choose this model as a straightforward extension from the $SU(2)$ case to a slightly more complicated group, however its actual physical implementation is not as clear. Some of the properties mentioned before still hold: the system is also fully controllable, and we show a proof of that in Appendix \ref{app:su3}. Also, the speed of evolution in state space is constant, since

\begin{eqnarray}
\frac{dS_1}{dt}&=&\Delta \eps = \Omega,\: \forall\: t \\
\frac{dS_2}{dt}&=&\frac{2}{\sqrt{d}}||H_3(t)||=\sqrt{\frac{2}{3}}\Omega,\: \forall\: t
\end{eqnarray}

Note that the two metrics are not equivalent for $d=3$. As we did before, we will introduce a family of unitaries that will serve as targets

\begin{equation}
V_X(\phi)=\ex{-i \lambda_X\phi},\ 0\leq \phi\leq \pi
\label{ec:targets_su3}
\end{equation}

\noindent where $-i\lambda_X$ are now elements of a basis of the Lie algebra, which are normalized such that $\trace{\lambda_X \lambda_Y}=2\delta_{X,Y}$. The elements of this basis are shown in Appendix \ref{app:su3}. Of particular interest here will be $\lambda_A$, $\lambda_B$ (already introduced) and

\begin{equation}
\lambda_C = -i\sqrt{\frac{2}{5}}\conmu{\lambda_A}{\lambda_B},\ \  \lambda_D=i\sqrt{\frac{4}{17}}\conmu{\lambda_A}{\lambda_C}
\end{equation}

\begin{figure}
	\includegraphics[width=0.9\linewidth]{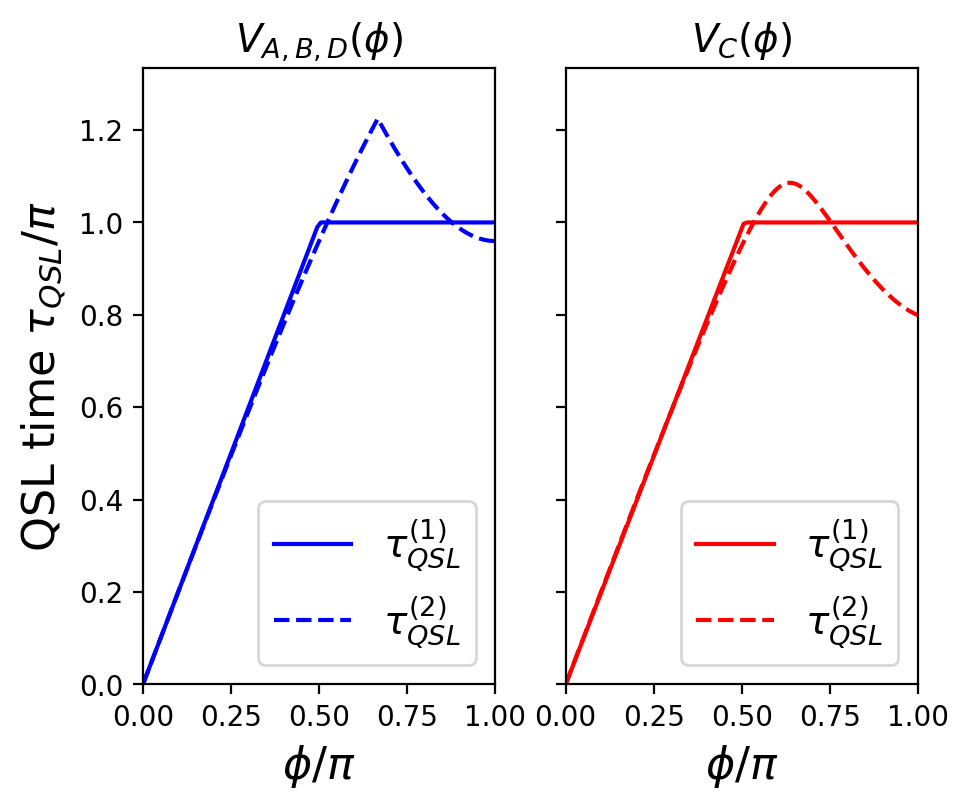}
	\caption{\label{fig:fig2} QSL times for the $SU(3)$ model described by the Hamiltonian (\ref{ec:hami_su3}) as a function of parameter $\phi$. Results are shown for different target unitaries $V_X$, defined in Eqn. (\ref{ec:targets_su3}). Left: $X=A,B,D$, which give the same QSL times. Right: $X=C$.  } 
\end{figure}

For this system, we also calculate the QSL bounds $\tau_{QSL}^{(i)}$ with $i=1,2$. Expressions are less transparent than in the $SU(2)$ case, so we show plots of them as a function of $\phi$ and $X$ in Fig. \ref{fig:fig2}. There, we can see that both quantities grow with $\phi$, as expected. Note also that picking the the bigger bound (i.e. the most meaningful) depends on which value of $\phi$ we are considering, and typically we define an unified bound which is simply $\tau_{QSL}=\max \{\tau^{(1)}_{QSL},\tau^{(2)}_{QSL}\}$.

\subsection{Side note: classical limit of the QSL}

Before we move on to analyzing how the QSL bounds relate to the minimum control time problem, let us explore another simple example which will further clarify the physical meaning of the quantum speed limit. As discussed in the previous section, the QSL time gives the minimum time that takes for the unitary to achieve a certain degree of distinguishability from identity. Since the first discussions on the QSL for states it has been noted that this notion was particular to quantum dynamics, and it can be seen that the Mandelstam - Tamm inequality (\ref{ec:qsl_st_mt}) becomes trivial when $\hbar\rightarrow 0 $. Recent works have shown that one could also derive speed limits for classical systems when the state is characterized by some distribution $\rho(x,p,t)$ in phase space \cite{delcampo2018,okuyama2018}. However, it is important to stress that this speed limit is not intrinsic to classical dynamics, since it only arises from ignorance about the state of our system. In quantum mechanics, a system will always need a finite amount of time to become completely distinguishable to its initial configuration. \\

Let us then explore the classical limit of the QSL formulation for unitary operations. We go back to considering $SU(2)$ rotations as in Eqn. (\ref{ec:hami_su2}), but we now choose the spin-$J$ representation, in which the generator takes the form

\begin{equation}
H(t) = \Omega \left(\cos{\alpha(t)}J_x + \sin{\alpha(t)}J_y\right),
\end{equation}  

\noindent where now $J_n$, with $n=x,y,z$ are the usual spin operators. Analogously to Eqn. (\ref{ec:targets_su2}) we define the target as $V_n(\phi)=\ex{-i J_n \phi}$. We can calculate the distance from $V_n$ to the identity in a straightforward way, yielding

\begin{equation}
S_2(V_n,\mathbb{I})=2\:\mathrm{arccos}\left(\left\lvert \frac{\sin\left((J+\frac{1}{2})\phi\right)}{(2J+1)\sin{\frac{\phi}{2}}}\right\rvert\right)
\end{equation}

The corresponding QSL time is then given by Eqn. (\ref{ec:qsl_un_t2}), where

\begin{equation}
||H(t)||^2 = \frac{J(J+1)(2J+1)}{3}\Omega^2.
\end{equation}

Note that maximum distinguishability is obtained when the unitary is orthogonal to identity, and this happens for a certain value of $\phi=\phi_\perp(J)$ (irrespective of the axis of rotation)

\begin{equation}
\phi_\perp(J)=\mathrm{argmax}_{\phi}\left[ S_2(V_n,\mathbb{I}) \right] = \frac{\pi}{J+\frac{1}{2}}.
\end{equation}

We observe that $\phi_\perp$ vanishes as $J\rightarrow\infty$, which is effectively the classical limit. In other words, the angle one needs to rotate to get an orthogonal unitary becomes smaller and smaller in this limit. Moreover, the rate at which the system evolves in state space is given by $||H(t)||$ which increases with $J$. Putting these results together it can be readily seen that

\begin{equation}
\tau^{(2)}_{QSL} \xrightarrow[J\rightarrow\infty]{}0
\end{equation}

It can be easily shown that this results is independent of the metric. In this way we can appreciate how the QSL time vanishes for classical systems.\\

\section{Minimum control time} \label{sec:control}

\subsection{Motivation}

In this Section we turn our attention to the minimum control time (MCT) problem and its relation to the QSL time. For that, let us recall the quantum control problem invoked in the Introduction. We consider a $d$-dimensional quantum system described by a traceless Hamiltonian $H[\epsilon(t)]$, where $\epsilon(t)$ denote the (typically) time-dependent control fields. The system evolves according to

\begin{equation}
i \frac{dU(t)}{dt}=H[\epsilon(t)]U(t),\ \mathrm{with}\ U(0)=\mathbb{I}
\label{ec:unit_schro}
\end{equation}

For a given target unitary transformation $V\in SU(d)$, we wish to study what is the minimum value of the evolution time $T$ for which this evolution can be such that $U(T)=V$. As mentioned before, obtaining the actual value of this minimum control time $T_{MCT}$ would require solving a time-optimal control problem. However, our aim is to obtain lower bounds to $T_{MCT}$, without resorting to any optimization procedure.\\

One of such bounds is, indeed, given by the geometric QSL formulation studied in Section \ref{sec:geom_qsl}, i.e. Eqns. (\ref{ec:qsl_un_t1}) and (\ref{ec:qsl_un_t2}). Note that these expressions give the same bound for all targets that are equidistant from the identity, and represent a fundamental limit to how fast the system can traverse that distance in space of unitaries. However, a more stringent limitation to how fast we can perform a transformation is given by the fact that we have limited control over the system. Formally, what this means is that only a small subset of the $d^2-1$ linearly independent generators can be directly manipulated. As we will discuss in the following, the relation between these directly-controlled generators and the target $V$ will have an important impact in the MCT.\\

For concreteness, consider that the Hamiltonian has the form
\begin{equation}
H(t) = \sum\limits_{j=1}^M \epsilon_j(t)H_j
\label{ec:hami_gral}
\end{equation}

To asses the controllability of the system, one introduces its \emph{dynamical Lie algebra} $\mathcal{L} = \mathrm{span}_{j=1,\ldots,M}\left\{-iH_j\right\}$. Note that the elements of this set will be given by the directly-controlled generators $\{H_j\}$ and also repeated commutators of them, i.e. elements of the form $\conmu{H_l}{H_m}$, $\conmu{H_l}{\conmu{H_m}{H_k}}$, etc. We define the \emph{depth} of each element of $\mathcal{L}$ as the maximum number of commutation operators required to generate it. If $\mathcal{L}=\mathfrak{su}(d)$, the system is said to be fully controllable \cite{dalessandro}.\\

Given a particular target $V=\ex{-i X \phi}$, we can qualitatively see that we can associate a depth in $\mathcal{L}$ to the target generator $X$. As noted previously in Ref. \cite{arenz2017}, we expect this depth to impact the MCT. In order to obtain some intuition about this, let us take as an example the two-level Hamiltonian of Eqn. (\ref{ec:hami_su2}) and the corresponding family of target gates of Eqn. (\ref{ec:targets_su2}). In this simple case, it is readily seen that targets of the form $V_x(\phi)$ and $V_y(\phi)$ are easily achieved by setting the field $\alpha(t)$ constant and evolving for a time $\phi/\Omega$, thus meaning that the QSL time coincides with the MCT for these cases. The attainability of this fundamental speed limit is a direct consequence of having direct control of the $\sigma_x$ and $\sigma_y$ in the Hamiltonian. On the other hand, the targets $V_z(\phi)$ require explicit time-dependence of the control fields and longer control times. In the following, we will explore this feature with a more systematic approach and analyze the short-time attainability of unitary transformations involving higher-depth elements of the dynamical Lie algebra. \\

\subsection{Nested commutators and analytic bounds for the MCT}

In order to make the ideas presented in the previous Subsection more concrete, let us rewrite the Schr\"odinger equation, c.f. Eqn. (\ref{ec:unit_schro}) in an alternative way. For that, we will use the formula for the derivative of the exponential map \cite{hall2015}. Given a Lie group $G$ and its associated Lie algebra $\mathfrak{g}$,  we can think of the exponential as a map $\mathrm{exp}:\mathfrak{g}:\rightarrow G$, such that if we have a $C^1$ path $X(t)$ in $\mathfrak{g}$ we can compute
\begin{equation}
\frac{d}{dt}\ex{X(t)}=\ex{X}\frac{1-\ex{-\ad{X}}}{\ad{X}}\frac{dX}{dt}.
\label{ec:ph_lemma}
\end{equation}

Here, $\ad{X}:\mathfrak{g}\rightarrow\mathfrak{g}$ is given by $\ad{X}(Y)=[X,Y]$. We will also use the definition of $\mathrm{Ad}_A(X)=AXA^{-1}$, where $A\in G$, and the important property that
\begin{equation}
\mathrm{Ad}_{\ex{x}}=\ex{\ad{X}},\ X\in\mathfrak{g}.\\
\end{equation}

Let us now write the evolution operator $U(t)=\ex{-iA(t)}\in SU(d)$ such that $-iA\in\mathfrak{su}(d)$. We can use (\ref{ec:ph_lemma}) to express Eqn. (\ref{ec:unit_schro}), which we will use as an starting point
\begin{equation}
\frac{dA}{dt}=\frac{\ad{-iA}}{\ex{\ad{-iA}}-1}H=\sum\limits_{m=0}^{\infty}\frac{B_m}{m!}(-i)^m\left(\ad{A}\right)^m H,
\label{ec:schro_ad}
\end{equation}

\noindent where $\{B_m\}$ are the set of Bernouilli numbers \cite{apostol2013}. Eqn. (\ref{ec:schro_ad}) is a differential equation for the generator $A(t)$ of $U(t)$ and will thus prove useful to analyze how different transformations become accessible at short times.\\

We will consider dimensionless units by introducing $H=\Omega h$ and $t=\frac{s}{\Omega}$. We then introduce a Taylor expansion of the Hamiltonian and the generator in the time variable
\begin{eqnarray}
A(s)&=&\sum\limits_{n=1} A^{(n)}s^n\\
h(s)&=&\sum\limits_{n=0} h^{(n)}s^n.
\end{eqnarray}

With this we can express Eqn. (\ref{ec:schro_ad}) as
\begin{equation}
\sum\limits_{n}nA^{(n)}s^{n-1}=\sum\limits_{m}(-1)^m\frac{B_m}{m!}\sum_k\left(\sum\limits_{n}s^n \ad{A^{(n)}}\right)^m h^{(k)}s^k.
\label{ec:schro_ad_2}
\end{equation}

Collecting powers of $s$ we get the following set of equations up to $\mathcal{O}(s^2)$
\begin{eqnarray}
A^{(1)}&=&h^{(0)}\\
2A^{(2)}&=&h^{(1)}-i\frac{B_1}{1!}\ad{A^{(1)}}h^{(0)}\\
3A^{(3)}&=&h^{(2)}-i\frac{B_1}{1!}\left(\ad{A^{(1)}}h^{(1)}+\ad{A^{(2)}}h^{(0)}\right) \nonumber \\
& &-\frac{B_2}{2!}\adc{A^{(1)}}h^{(0)}
\end{eqnarray}

These equations can be solved iteratively and give
\begin{eqnarray}
A^{(1)}&=&h^{(0)} \label{ec:A1} \\
A^{(2)}&=&\frac{1}{2}h^{(1)} \label{ec:A2}\\
A^{(3)}&=&\frac{1}{3}h^{(2)}+\frac{i}{12}[h^{(0)},h^{(1)}] \label{ec:A3}
\end{eqnarray} 

In Appendix \ref{app:pert} we extend this analysis up to $\mathcal{O}(s^4)$. Note that Eqn. (\ref{ec:A3}) involves the commutator of different terms in the Hamiltonian, which will determine the appearance of depth-1 elements of the dynamical Lie algebra in the evolution of $A(t)$, but only in terms proportional to $s^3$ or higher.\\

In order to analyze how restricting the available terms in the Hamiltonian is reflected in this equations, we introduce a set $\left\{\chi_\mu \right\}$, $\mu=1,\ldots,d^2-1$ such that $\{-i \chi_\mu \}$ is an orthogonal basis of $\mathfrak{su}(d)$, normalized such that $\trace{\chi_\mu\chi_\nu}=\delta_{\mu \nu}$.  We can then expand

\begin{eqnarray}
A^{(n)}(s)&=&\sum\limits_{\mu=1}^{d^2-1} a^{(n)}_\mu(s) \chi_\mu\\
h^{(n)}(s)&=&\sum\limits_{\mu=1}^M \epsilon^{(n)}_\mu(s)\chi_\mu,
\end{eqnarray}

Here, the key point is that the number of elements in the expansion of $h(s)$, $M$, is typically much smaller than $d^2-1$. From now on we will specialize in the simplest non-trivial case, where we have $M=2$ orthogonal terms, to develop bounds and estimates on the MCT. Generalizations to higher number of terms can be pursued with the same formalism. Thus we can write 

\begin{equation}
h=\epsilon_A(s) \chi_A + \epsilon_B(s) \chi_B,
\label{ec:hami_m2}
\end{equation}

\noindent and also define

\begin{equation}
\chi_C = \frac{-i}{f_{ABC}}\conmu{\chi_A}{\chi_B}
\end{equation}

 \noindent where $f_{ABC}$ is the corresponding structure constant of the group. Note that $\chi_C$ is the only depth-1 element in the algebra (to avoid confussion we are using capital letters $A,B,\ldots$ for the values of the greek-letter indices $\mu,\nu$), and that $\{\chi_A,\chi_B,\chi_C\}$ form an orthogonal set. We can then write the explicit form of the short-time solution from Eqns. (\ref{ec:A1})-(\ref{ec:A3}).

\begin{eqnarray}
a_A(s) &=& \epsilon_A^{(0)}s+\frac{1}{2}\epsilon_A^{(1)}s^2+\frac{1}{3}\epsilon_A^{(2)}s^3 \label{ec:aA_sol}\\
a_C(s) &=& -\frac{f_{ABC}}{12}\left(\epsilon_A^{(0)}\epsilon_B^{(1)}-\epsilon_A^{(1)}\epsilon_B^{(0)}\right)s^3, \label{ec:aCsol}
\end{eqnarray}

\noindent which are valid up to $\mathcal{O}(s^3)$ (the solution for $a_B$ is identycal to the one for $a_A$). From these its easy to derive bounds for the evolution time. Let us first assume the case of phase control, where $\epsilon_A(s)=E \cos{\alpha(s)}$ and $\epsilon_B(s)=E \sin{\alpha(s)}$. Then, from Eqn. (\ref{ec:aA_sol}) it follows that the time $s_A$ needed to achieve $a_A(s_A)=\beta$ obeys
\begin{equation}
s_A\geq \frac{\beta}{E}.
\label{ec:tau_A}
\end{equation}

Similarly, the time $s_C$ for which $a_C(s_C)=\beta$ follows the inequality

\begin{equation}
s_C \geq \sqrt{\frac{12 \beta}{f_{ABC}E^2}}
\label{ec:tau_C}
\end{equation}

Finally, in Appendix \ref{app:pert} we derive an analogous inequality associated to a depth-2 element of the algebra $\chi_D$,

\begin{equation}
s_D \geq \left(\frac{18 \beta}{f_{ABC}f_{ACD} E^3}\right)^{\frac{1}{3}}.
\label{ec:tau_D}
\end{equation}

These inequalities are the main analytical results of this paper. They are valid for short times (i.e. $s\ll 1$) and they apply to any Hamiltonian of the form (\ref{ec:hami_m2}).To get a feel of the information these bounds give on the minimum control time, we will apply them to the two model systems introduced in Section \ref{sec:geom_qsl}. For $SU(2)$ we identify $A=x$ and $C=z$ to obtain

\begin{eqnarray}
T_x &\geq& \frac{\phi}{\Omega} \equiv \tau_x(\phi) \label{ec:tau_A_su2}\\
T_z  &\geq& \frac{\sqrt{12 \phi}}{\Omega} \equiv \tau_z(\phi) \label{ec:tau_C_su2}
\end{eqnarray}

These two bounds are a refinement of the geometrical QSL bound of Eqn. (\ref{ec:qsl_t_su2}). Note that $\tau_x=\tau_{QSL}$, as expected from the previous discussion, and that $\tau_z > \tau_{QSL}$. In this way, the new bounds are more restrictive than the QSL, and thus are more meaningful to bound the MCT. This is achieved thanks to the fact that we have used information about the available controls in the Hamiltonian. It is important also to remark that we have not solved any optimization to get to these results: the expressions obtained are completely independent of the actual shape of the the control field $\alpha(s)$.\\

For the $SU(3)$ model we will proceed to compare these bounds directly with the numerical results for the MCT in the next section.\\

\subsection{Numerical results for the MCT}

In this Section we compare the analytical bounds of Eqns. (\ref{ec:tau_A})-(\ref{ec:tau_D}) with numerical results  of the MCT for both models discussed throughout this paper. First, we briefly describe the methodology to estimate (or rather, upper bound) the MCT, which has been applied to a wide variety of systems in several works over the past decade. Examples include Landau-Zener models \cite{caneva2009,poggi2015}, one-dimensional chains of interacting spin-$\frac{1}{2}$ particles \cite{caneva2011}, entangling transformations in superconducting qubits \cite{egger2013} and neutral atoms \cite{goerz2011}.\\

\begin{figure}[!t]
	\includegraphics[width=0.9\linewidth]{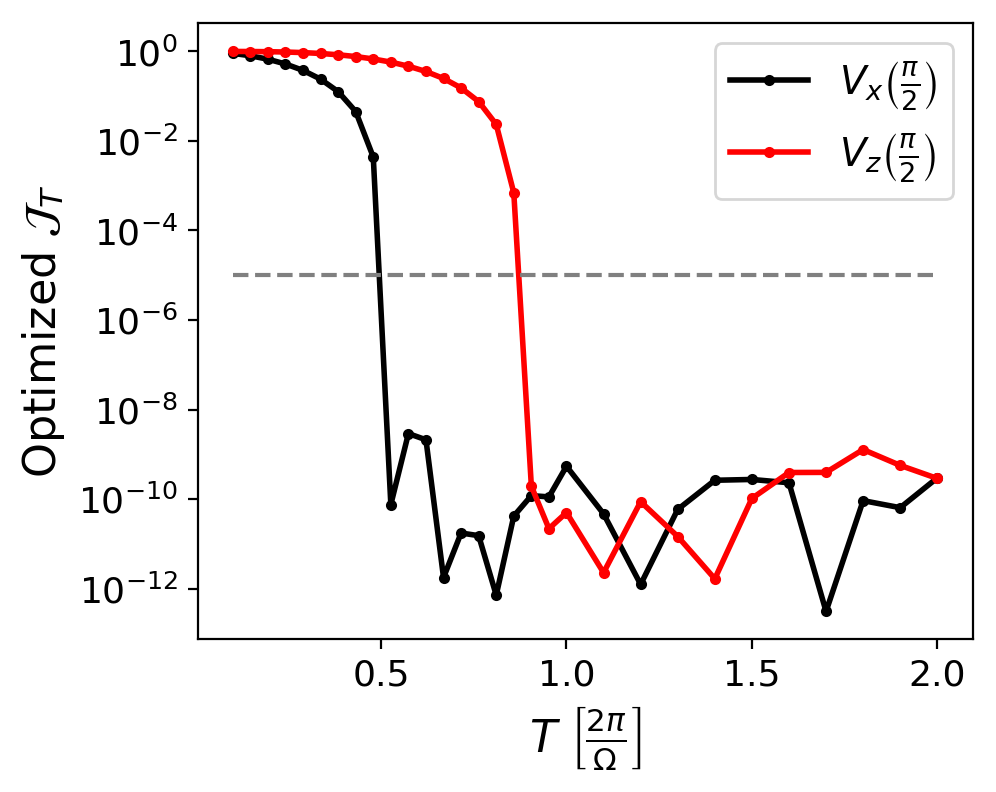}
	\caption{\label{fig:fig3} Numerical calculation of the MCT. We show the optimized functional value as a function of the total evolution time $T$ given to the optimal control algorithm. The MCT is estimated here as the minimum value of $T$ for which the infidelity $\mathcal{J}_T$ reaches $10^{-5}$. Results are shown for a specific example, given by the $SU(2)$ model with the target parameter $\phi=\frac{\pi}{2}$. The number of time steps is set to $N_{ts}=30$, and the results do not change significantly upon increasing this value. }
\end{figure}

The procedure is as follows. Given the Hamiltonian $H[\epsilon(t)]$ and a target unitary transformation $V$, we fix the total evolution time $T$ and define an optimization functional 
\begin{equation}
\mathcal{J}_T = 1-\frac{1}{d^2}\lvert \trace{\daga{V}U(T)}\rvert^2,
\end{equation}

\noindent which is the infidelity between the target and $U(T)$, the unitary evolution operator generated by $H(t)$ at $t=T$. For each $T$ and starting from a random initial control field $\epsilon^{(0)}(t)$ we use the GRAPE algorithm \cite{khaneja2005,motzoi2011} to find the optimum field $\epsilon^{*}(t)$ that minimizes $\mathcal{J}_T$. In practice, the time variable is divided into $N_{ts}$ steps, such that the optimization is carried over the $N_{ts}$ variables that parameterize the piecewise-constant control field. Starting from a suitably large value of $T$, once the optimum is found, that solution is fed as an initial seed for the next step, where $T$ is now reduced. This procedure is repeated for several random initial fields and, in each case, the best attained infidelity is recorded. A typical set of results arising from this procedure is shown in Fig. \ref{fig:fig3}. We then calculate $T_{min}$ as the minimum value of $T$ such that the functional achieves a certain threshold, which is set to $10^{-5}$ throughout this paper. Notice that the optimized $\mathcal{J}_T$ follows a sharp transition at $T=T_{min}$, and so the the results are quite insensitive to the choice of threshold.\\

\begin{figure}[!t]
	\includegraphics[width=0.9\linewidth]{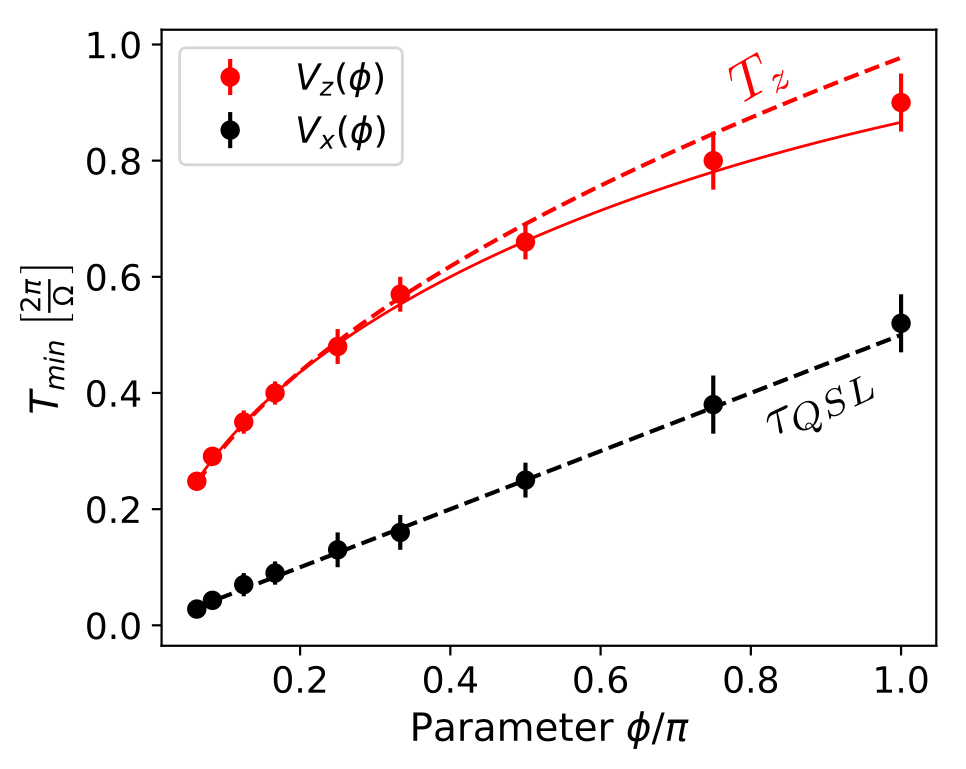}
	\caption{\label{fig:fig4} Minimum control time (MCT) for the $SU(2)$ model. Results are shown for the targets $V_x(\phi)$ and $V_z(\phi)$ and compared with the different bounds studied in this paper. Black dashed lines shows the geometric QSL time, while the red dashed line shows the short-time bound $T_z$ derived in Section \ref{sec:control}. Full red line shows the analytical result for the MCT for this particular model, derived in Ref. \cite{boozer2012}. }
\end{figure}

Using the procedure outlined in the above paragraph, we have numerically calculated the MCT for several cases of interest. In Fig. \ref{fig:fig4} we show results corresponding to $SU(2)$ for the family of targets in Eqn. (\ref{ec:targets_su2}), specifically $V_x(\phi)$ (which are identical to those of $V_y$) and $V_z(\phi)$. There, we can see that minimum time to achieve $z$-rotations is considerably bigger than the minimum time required to perform $x$-rotations, as expected. The dependence with the angle of rotation $\phi$, which determines how far away the target transformation is from the identity, is also different for both cases.\\

Together with the numerical results, in Fig. \ref{fig:fig4} we have plotted the geometric QSL bound of Eqn. (\ref{ec:qsl_t_su2}), and the short-time bounds $T_z$ in Eqn. (\ref{ec:tau_C_su2}) and $T_x$ (which coincides with $\tau_{QSL}$ here). We can verify here that the MCT for $V_x$ saturates the QSL bound. The MCT for $V_z$, on the other hand, is considerably bigger than the QSL time, but it matches the short-time bound we have proposed. Note that the MCT in this case actually surpasses $T_z$ for targets that further away from identity. This is reasonable since the inequality (\ref{ec:tau_A_su2}) is strictly valid for $\Omega t\ll 1$. However, $T_z$ estimates the actual MCT much more accurately than the QSL. This property is useful, since it allows to provide a higher-level estimation to the MCT \emph{before} solving any intrincate time-optimization problem. Note that, for this particular model, an actual analytical calculation for the MCT was proposed and solved in Ref. \cite{boozer2012}. The result, which roughly coincides with the numerical estimates, is also shown in the plot.\\

\begin{figure}[!t]
	\includegraphics[width=0.9\linewidth]{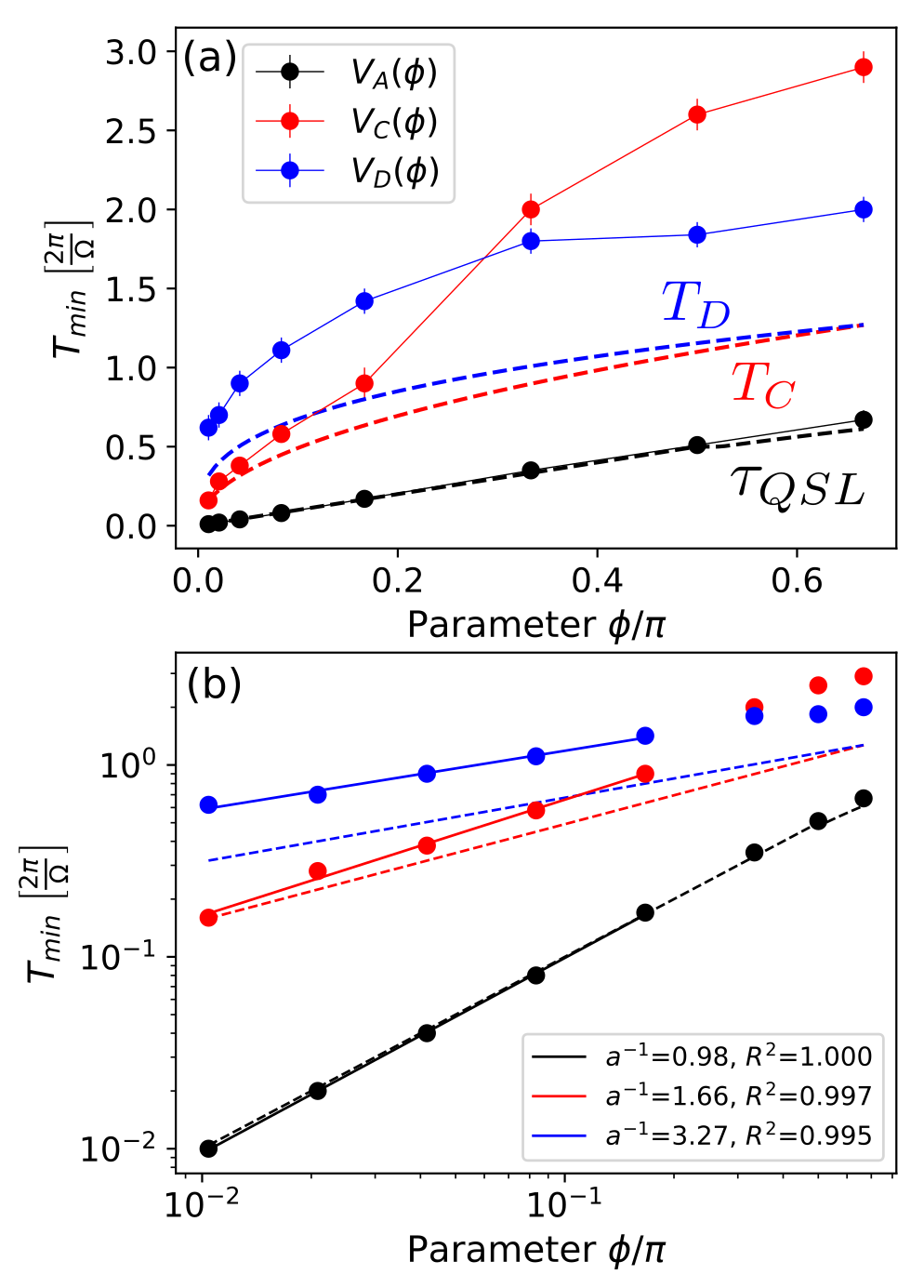}
	\caption{\label{fig:fig5} Minimum control time (MCT) for the $SU(3)$ model. Results are shown for the targets $V_A(\phi)$, $V_C(\phi)$ and $V_D(\phi)$ and compared with the different bounds studied in this paper. (a) Black dashed lines shows the geometric QSL time, while the red and blue dashed lines shows the short-time bounds $T_C$ and $T_D$ derived in Section \ref{sec:control}. (b) Same data as in (a) displayed in a log-log plot. Thin full lines indicate a power law fit of the form $y=bx^{a}$. The inverse powers obtained are displayed in the legend, together with the correlation coefficient $R^2$. }
\end{figure}

We now turn our attention to the $SU(3)$ model. Results for the MCT are shown in Fig. \ref{fig:fig5} (a). Here we have analyzed targets corresponding to depths 0 ($V_A$), 1 ($V_C$) and 2 ($V_D$) in the dynamical Lie algebra of the Hamiltonian (\ref{ec:hami_su3}). Comparing to the $SU(2)$ example, the situation is now considerably more complicated. Interestingly, we note that the depth does not imply an hierarchy in the MCT, i.e., for large enough $\phi$, the target $V_D(\phi)$ is achievable in a shorter time than $V_C(\phi)$, albeit the former being associated with a higher depth. Nevertheless, when analyzing targets close to identity, i.e. when $\phi\ll 1$ we do observe such hierarchy which is well predicted by our analysis in the beginning of this section.\\

Fig. \ref{fig:fig5} (a) also shows the relevant bounds computed for this problem. Note that, apart from $\tau_{QSL}=T_A$ (which is the simplest case), the other bounds are not saturated. This is expected in the general case, since the aim of this analysis is to bound or estimate the MCT, and obtaining its precise value implies solving an optimization. Nonetheless, the short-time bounds $T_C$ and $T_D$ do give valuable information: first, they are much better bounds than the QSL time. From Fig. \ref{fig:fig5} (b), which shows the same data as (a) but in a log-log plot, we can see that both of them predict very well the order of magnitude of the MCT. Note that this is very relevant, since the geometric QSL time can be almost two orders of magnitude below the MCT, as in the case of $V_D$. The short-time bounds also predict nicely the power law behaviour of the MCT for small $\phi$, the power being the inverse of associated depth of the target plus one. \\

\section{Summary and outlook} \label{sec:conclusions}

In this paper we have explored the relation between the geometric quantum speed limit and the minimum control time for unitary transformations in finite-dimensional quantum systems. In order to do this, we have first taken the QSL formalism for pure-state evolution and constructed inequalities that bound the evolution time of $U(t)$, the unitary evolution operator. While these inequalities are universally valid, we point out that, in the context of quantum control problems, more stringent limitations arise due to limited control over the system. Following this key observation, we systematically analyzed the short-time behaviour of the time-dependent Schr\"odinger equation, which allowed us to derive a new family of bounds that explicitly show the role of the so-called Dynamical Lie Algebra on the minimum control time. We have applied our results to $SU(2)$ and $SU(3)$ models and show how, even in low-dimensional systems, the proposed bounds can give much more information on the MCT than the geometric QSL.\\

We point out that the bounds proposed in Sect. \ref{sec:control} are general, although its explicit calculation for higher-dimensional systems and higher-order depths will become intricate. Nevertheless, we expect that they would give much better estimates for the MCT than the QSL in any case. These estimates on the MCT are important since they give a baseline over which we can use quantum optimal control to refine results.\\

Finally we wish to point out some interesting potential generalizations of this work. First, it is important to asses the role of drift terms in the Hamiltonian, i.e. terms over which we have no control and are typically constant in time. These is the usual case for two-qubit systems in most quantum computing platforms, where the two-body (entangling) interaction is usually fixed by the specifications of the device. It would be interesting to address this in connection to the Zermelo navigation problem \cite{russell2014,brody2015}.  Second, the short-time bounds presented here are derived from a perturbative analysis, but it would be interesting to study if we can develop a theoretical description that goes beyond the QSL and connects with the time-optimal control problem using the tools of differential geometry \cite{wang2015,nielsen2006}. 

\begin{acknowledgments}
The author gratefully acknowledges insightful discussions with Chris S. Jackson, Anupam Mitra and Ivan Deutsch. This work was partially supported by National Science Foundation Grant No. PHY-1630114
\end{acknowledgments}

\bibliography{qsl_unitaries}

\appendix

\section{Controllability of the $SU(3)$ model} \label{app:su3}

Here we explicitly show that the $SU(3)$ model of Eqn. (\ref{ec:hami_su3}) is fully controllable. As discussed in Sect. \ref{sec:control}, to prove this we have to show that its dynamical Lie algebra $\mathcal{L}$ equals $\mathfrak{su}(3)$. Following the procedure outlined in \cite{dalessandro}, we start from the depth-0 elements of $\mathcal{L}$, namely $\lambda_A=\lambda_1$, $\lambda_B=(\lambda_2+\lambda_4)/\sqrt{2}$ (already introduced in the main text), we calculate all the possible nested commutators and check that we can form a set of $d^2-1 = 8$ linearly independent elements.\\

The only depth-1 element will be related to the commutator of these two operators,
\begin{equation}
\conmu{\lambda_A}{\lambda_B}\propto \lambda_C= \sqrt{\frac{4}{5}}\left(\lambda_3 + \frac{1}{2}\lambda_7\right).
\end{equation}

Depth-2 elements are given by
\begin{eqnarray}
\conmu{\lambda_A}{\conmu{\lambda_A}{\lambda_B}}&\propto& \lambda_D=\frac{4}{\sqrt{17}}\left(\lambda_2 + \frac{1}{4}\lambda_4\right)\\
\conmu{\lambda_B}{\conmu{\lambda_A}{\lambda_B}}&\propto& \lambda_E=-\frac{5\sqrt{2}}{2\sqrt{17}}\left(\lambda_1 + \frac{3}{5}\lambda_5\right) \nonumber
\end{eqnarray}

Finally, the depth-3 elements are
\begin{eqnarray}
\conmu{\lambda_A}{\conmu{\lambda_A}{\conmu{\lambda_A}{\lambda_B}}}&\propto& \lambda_F=\frac{4\sqrt{4}}{\sqrt{65}}\left(\lambda_3 + \frac{1}{8}\lambda_7\right) \\
\conmu{\lambda_A}{\conmu{\lambda_B}{\conmu{\lambda_A}{\lambda_B}}}&\propto& \lambda_G=\lambda_6 \nonumber \\
\conmu{\lambda_B}{\conmu{\lambda_A}{\conmu{\lambda_A}{\lambda_B}}}&\propto& \lambda_G=\lambda_6 \nonumber \\
\conmu{\lambda_B}{\conmu{\lambda_B}{\conmu{\lambda_A}{\lambda_B}}}&\propto& \lambda_H=\frac{1}{\sqrt{65}}\left(\frac{13}{2}\lambda_3 + 4\lambda_7+\frac{3\sqrt{3}}{2}\lambda_8\right) \nonumber
\end{eqnarray}

It can be readily verified that the set of eight operators $\left\{\lambda_X \right\} \in \mathcal{L}$ with $X=A,B,\ldots,H$ are linearly independent. Since all of these are elements of $\mathfrak{su}(3)$, then it follows that $\mathcal{L}=\mathfrak{su}(3)$, meaning that the system is fully controllable. 

\section{Perturbative solution to Schr\"odinger equation} \label{app:pert}

Continuing to collect powers of $s$ in Eqn. (\ref{ec:schro_ad_2}) up to $s^4$ we obtain the following solutions
\begin{eqnarray}
A^{(4)}&=&\frac{1}{4}h^{(3)}+\frac{i}{12}[h^{(0)},h^{(2)}] \label{ec:A4}\\
A^{(5)}&=&\frac{1}{5}h^{(4)}+\frac{i}{20}\left(\frac{3}{2}[h^{(0)},h^{(3)}]+\frac{1}{3}[h^{(0)},h^{(2)}]\right) \nonumber \\
& &+\frac{1}{120}\left(\frac{1}{2}[h^{(1)},[h^{(0)},h^{(1)}]]-\frac{1}{3}[h^{(0)},[h^{(0)},h^{(2)}]]\right) \nonumber \\
& & +\frac{i}{720}[h^{(0)},[h^{(0)},[h^{(0)},h^{(1)}]]] \nonumber \label{ec:A5}
\end{eqnarray}

For the particular case discussed in the main text, where the Hamiltonian has $M=2$ terms coupling to the operators $\chi_A$ and $\chi_B$, we can define $\chi_D=\frac{-i}{f_{ACD}}\conmu{\chi_A}{\chi_C}$. If $\trace{\chi_D^2}=1$ and assuming that now the set $\left\{\chi_A,\chi_B,\chi_C,\chi_D \right\}$ is orthonormal, we get an equation for $a_D(s)$
\begin{equation}
a_D(s)=\frac{f_{ABC} f_{ACD}}{120} F(\epsilon) s^3 + \mathcal{O}(s^4),
\label{ec:aD}
\end{equation}
\noindent where
\begin{eqnarray}
F(\epsilon)&=&\epsilon_A^{(0)}\left(\frac{1}{2}\epsilon_A^{(1)}s \epsilon_B^{(1)}s - \frac{1}{3}\epsilon_A^{(0)} \epsilon_B^{(2)}s^2\right) \nonumber \\
& & - \epsilon_B^{(0)}\left(\frac{1}{2}\epsilon_A^{(1)}s \epsilon_A^{(1)}s - \frac{1}{3}\epsilon_A^{(0)} \epsilon_A^{(2)}s^2\right)
\end{eqnarray}

Following the phase control models discussed in this work, we take $\epsilon_A(s)=E \cos{\alpha(s)}$ and $\epsilon_B(s)=E \sin{\alpha(s)}$. Then, it can be shown that 
\begin{equation}
|F(\epsilon)|\leq \frac{20}{3}E^3.
\label{ec:cota_F}
\end{equation}

Combining Eqns. (\ref{ec:aD}) and (\ref{ec:cota_F}), we can see that the time $s_D$ for which $a_D(s_D)=\beta$ follows the inequality
\begin{equation}
s_D \geq \left(\frac{18 \beta}{f_{ABC}f_{ACD} E^3}\right)^{\frac{1}{3}},
\end{equation}

\noindent which is Eqn. (\ref{ec:tau_D}) in the main text. \\

Finally, note that if the two generators appearing in the Hamiltonian are orthogonal, then we can always associate the depth-0 and -1 elements of $\mathcal{L}$ to elements of an orthogonal basis. However, this is not generally true for depths higher than one. That is actually the case in the $SU(3)$ model, where $\lambda_D$ is not orthogonal to $\lambda_B$. Nevertheless, we can always generate an orthogonal basis by using the Gram-Schmidt process. If $\chi_D$ is the orthonormalized element corresponding to $\lambda_D\propto \tilde{\chi}_D$, then an estimate on the minimum time such that $\tilde{a}_D(s_D)=\beta$ is obtained by replacing $f_{ACD}$ with $\eta_D f_{ACD}$, where $\eta_D=\trace{\chi_D \tilde{\chi}_D}$.\\ 

\end{document}